\documentclass[useAMS,usenatbib]{mn2e}
\usepackage{graphicx}
\usepackage{lscape}

\title[\textsc{[OIII]}$\lambda 4363$ \AA\ emitting galaxies at intermediate $z$.]{Direct determination of 
oxygen abundances in line emitting star-forming galaxies at intermediate redshift.}

\author[J.~M.~P\'erez et al.]
{Jos\'e~M.~P\'erez,$^{1,2,3}$\thanks{E-mail: jm.perez@univie.ac.at}
Carlos~Hoyos,$^{1,3}$ \'Angeles~I.~D\'{\i}az,$^{1,3}$ David~C.~Koo,$^{4}$
\newauthor and Christopher~N.~A~Willmer$^{5}$ \\
$^{1}$Departamento de F\'{\i}sica Te\'orica. Universidad Aut\'onoma de Madrid, Spain.\\
$^{2}$Department of Astrophysics, University of Vienna, T\"urkenschanzstrasse 17,1180 Wien, Austria.\\
$^{3}$Astro-UAM, UAM, Unidad Asociada CSIC.\\
$^{4}$Department of Astronomy and Lick Observatory, University of California, 1156 High Street, Santa Cruz, CA 95064, USA\\
$^{5}$Department of Astronomy and Steward Observatory, University of Arizona, 933 North Cherry Avenue, Tucson, AZ 85721-0065, USA\\}

\begin{document}

\date{\today}
\pagerange{\pageref{firstpage}--\pageref{lastpage}} \pubyear{2015}
\maketitle
\label{firstpage}

\small
\begin{abstract}
We present a sample of 22 blue ($(B-V)_{AB}<0.45$), luminous ($M_{B,AB}<-18.9$), metal-poor
galaxies in the $0.69<z<0.88$ redshift range, selected from the DEEP2 galaxy redshift survey.
Their spectra contain the [\textsc{OIII}] $\lambda4363$
auroral line, the [\textsc{OII}]$\lambda \lambda3726,3729$ doublet
and the strong nebular [\textsc{OIII}]$\lambda \lambda 4959,5007$ emission lines.
The ionised gas-phase oxygen abundances of these
galaxies lie between $7.62<12+\log O/H < 8.19$, i.e. between $1/10 Z_{\odot}$ and $1/3 Z_{\odot}$.
We find that galaxies in our sample have comparable metallicities
to other intermediate-redshift samples, but are more metal poor
than local systems of similar B-band luminosities and star formation
activity. The galaxies here show similar properties to the ``green peas'' discovered at
$z\simeq 0.2 - 0.3$  though our galaxies tend to be slightly less
luminous.

\end{abstract}

\begin{keywords}
galaxies -- abundances -- evolution -- high-redshift.
\end{keywords}

\normalsize

\section{Introduction}

The study of metal poor, compact star forming galaxies was initiated
by \citet{sarsearl72}. Their importance was rapidly recognized because they were
considered to be ideal benchmarks in the study of the earliest stages of galaxy
evolution, and were studied in many works
(\citet{campbell88,pagel92,french80}, among many others).
While these nearby systems were originally thought to be genuinely young
systems, experiencing their very first star forming episodes, even the
most metal poor objects in the local universe have been found to posses
significant underlying evolved stellar populations,
 e.g., \citep{papa96,taylor95}. In fact, very deep Hubble Space Telescope
images of the most metal-poor object known, I Zw 18, indicate that it has formed
the bulk of its present stellar population 0.5--1.0 Gyr ago \citep{aloisitosi99}.
Other works on the stellar populations of this and similar galaxies yield
similar results \citep{papa02,aloisi07}. A motivation to 
examine samples at higher redshifts is to locate truly young systems
that could be experiencing their first major
star formation episodes. Some exploratory steps in this direction were taken by
\citet{hoyoskoo05}, who found that distant ($z \sim 0.7$), compact star
forming galaxies showing the [\textsc{OIII}]$\lambda$4363 line deviated from the
usual Luminosity-Metallicity relationships, implying that these distant, metal
poor systems might turn out to be truly young systems. One of the aims of this paper is
to further explore this hypothesis.

Knowledge of the metallicity of external galaxies is crucial for galaxy evolution theories
because it is a direct result of the integral history of the star formation
and mass assembly of galaxies. The metallicity of galaxies has been traditionally
used in conjunction with the luminosity to build the luminosity-metallicity relation (LZR)
see, e.g., \citet{lequeux79,rich_call95}. This relationship is thought to arise from
depth of the gravitational well of massive galaxies which does not
allow newly created metals being  cast into the ISM amidst
hot supernovae ejecta, where it is assumed that the more luminous galaxies are also the most
massive, and have a higher star formation efficiency. In this model,
lower mass systems will not be able to keep their heavy elements and
will probably be 
of lower metallicity. Galaxies in cluster environments undergo other processes, which greatly complicate the picture.
All these phenomena will naturally create a complex metallicity distribution for galaxies at various redshifts, of which only a few data points
at intermediate redshifts are precisely known.

This paper addresses the star formation history of galaxies by
examining the properties of a sample of star forming galaxies
showing the [\textsc{OIII}]$\lambda 4363$ line.

The [\textsc{OIII}]$\lambda$4363 auroral line is generated by the
collisionally excited transition (2p$^{2 1}$D to 2p$^{2 1}$S) 
of doubly ionised oxygen atoms. This line is key in the study of the metallicity of the ionised phase
of the interstellar medium (ISM) since it allows determining the electron temperature of ionised regions
without any previous assumption on the metal content of the observed nebula \citep{ost89}.
Unfortunately, the [\textsc{OIII}]$\lambda$4363 line is difficult to
detect for several reasons:
(i) This line is stronger in low metallicity clouds. In higher metallicity environments, it is
exponentially depressed because the increased cooling leaves no energy to excite oxygen atoms to the upper level
of this transition. (ii) This line is also stronger in young and strong starbursts. Older starbursts (older than 5--6 Myr)
are unable to maintain a large fraction of oxygen atoms doubly ionised. (iii) This line is also stronger in systems in which the
contribution to the continuum from the underlying stellar population is less important relative to the contribution
from the newly created ionising stars. The latter case is the one where
this weak line is less likely to be obliterated by 
the continuum noise. In \citet{hoyosdiaz06}, it was shown that all these issues affect the detection
of the [\textsc{OIII}]$\lambda$4363 \AA\ line for the case of local \textsc{HII} galaxies.

The discussion above explains why the majority of the most accurate metallicity determinations for intermediate redshift sources
have been obtained using measurements of the auroral oxygen lines in galaxies with strong starbursts. At the same time, it also explains
why these measurements may not be fully tracing the complete metallicity distribution at each redshift, since the upper end of this distribution is increasingly difficult
to trace accurately through the use of the oxygen lines. Furthermore,
even  in the $12+\log O/H \sim 8.10-8.35$ regime, the oxygen-based
strong line calibrators such as $\mathrm{R}_{23}$ allow two solutions, a problem difficult to overcome.
An alternative is to use  [\textsc{SIII}]$\lambda \lambda 9096, 9532$,
suggested for instance by \citet{2006A&A...449..193P}, which does
not exhibit the degeneracies that $\mathrm{R}_{23}$ has at intermediate metallicities. Finally, it is also possible 
to determine electron temperatures through the detection of [\textsc{SIII}]$\lambda 6312$ line up to at least 
solar metallicities and then derive good abundances for high-metallicity, vigorously star forming systems.
This has been shown in \citet{1994A&A...282L..37K}, \citet{2005A&A...441..981B}, \citet{2000MNRAS.318..462D}, and \citet{2002MNRAS.329..315C}.
One disadvantage in the use of sulfur lines is that they shift
into the NIR for objects at even moderate redshifts. However, the work
of \citet{loschinos2013} shows that this is being addressed with new
IR instrumentation. The latter work presents a study on the nature of
the [\textsc{SIII}]$\lambda \lambda 9096, 9532$ emitters,
showing that these objects are usually star forming systems where a Compton-thick AGN has little or no effect in exciting the sulfur lines. 

The work presented in \cite{hoyoskoo05} also investigated the differences between star forming systems with and without the [\textsc{OIII}]$\lambda$4363 \AA\ line
using intermediate redshift ($z\simeq0.7$) galaxies observed by the DEEP2 survey. Compact, star forming
galaxies showing the [\textsc{OIII}]$\lambda 4363$ auroral line have lower metallicities and higher emission
line equivalent widths than objects without this feature, in spite of their
H$\beta$ line luminosities  not being lower. This was also found by \citet{hoyosdiaz06}.
The underlying stellar populations of galaxies presenting the [\textsc{OIII}]$\lambda$4363 are
less luminous relative to their newly created ionising populations,
when compared to other star forming galaxies \citep{hoyosdiaz06}.

In this paper, \S \ref{ods} describes the observations and sample
selection, \S \ref{resultaos} shows our metallicity calculations based
on electron temperatures.We compare our results to other recent works dealing with both local and intermediate
redshift sources in \S \ref{discutir}. This section also 
summarizes the paper. We use $(\Omega_{M},\Omega_{\Lambda},h_{70})=(0.3,0.7,1)$. Magnitudes are 
given in the $AB$ system. 

\section{Observations and Sample Selection.}
\label{ods}

The data used in this work were taken for the second phase of the Deep Extragalactic Evolutionary Probe survey
(DEEP2\footnote{See \texttt{http://deep.ucolick.org/}}, \citet{deep2}).
This survey uses the DEIMOS\footnote{See \texttt{http://www2.keck.hawaii.edu/inst/deimos/}} \citep{deimospaper} spectrograph on the W.M. Keck telescope.
DEEP2 is a densely sampled, high precision redshift survey which, thanks to a
$BRI$ colour pre-selection criterion, preferentially targets galaxies in
the $0.7<z<1.4$ redshift range in three of the four areas of the sky
it covers. 
It has collected a grand total of 53000 spectra, measuring 38000 reliable redshifts.
The DEIMOS instrument was used with a 1200$\mathrm{mm}^{-1}$ grating
centered at 7800\AA, thus covering on average the 6500\AA-9100\AA \ 
wavelength range. The wavelength range shows slight changes that
depend on the slit position on the mask. The
resolving power ($R \simeq 5000$) allows separating the [\textsc{OII}]$\lambda\lambda3726,3729$ doublet.
Because of the relatively high resolution, the DEEP2 spectra also
yield accurate velocity dispersions for emission line objects.

\begin{figure*}
\begin{center}

\begin{tabular}{cc}
\includegraphics[scale=0.50]{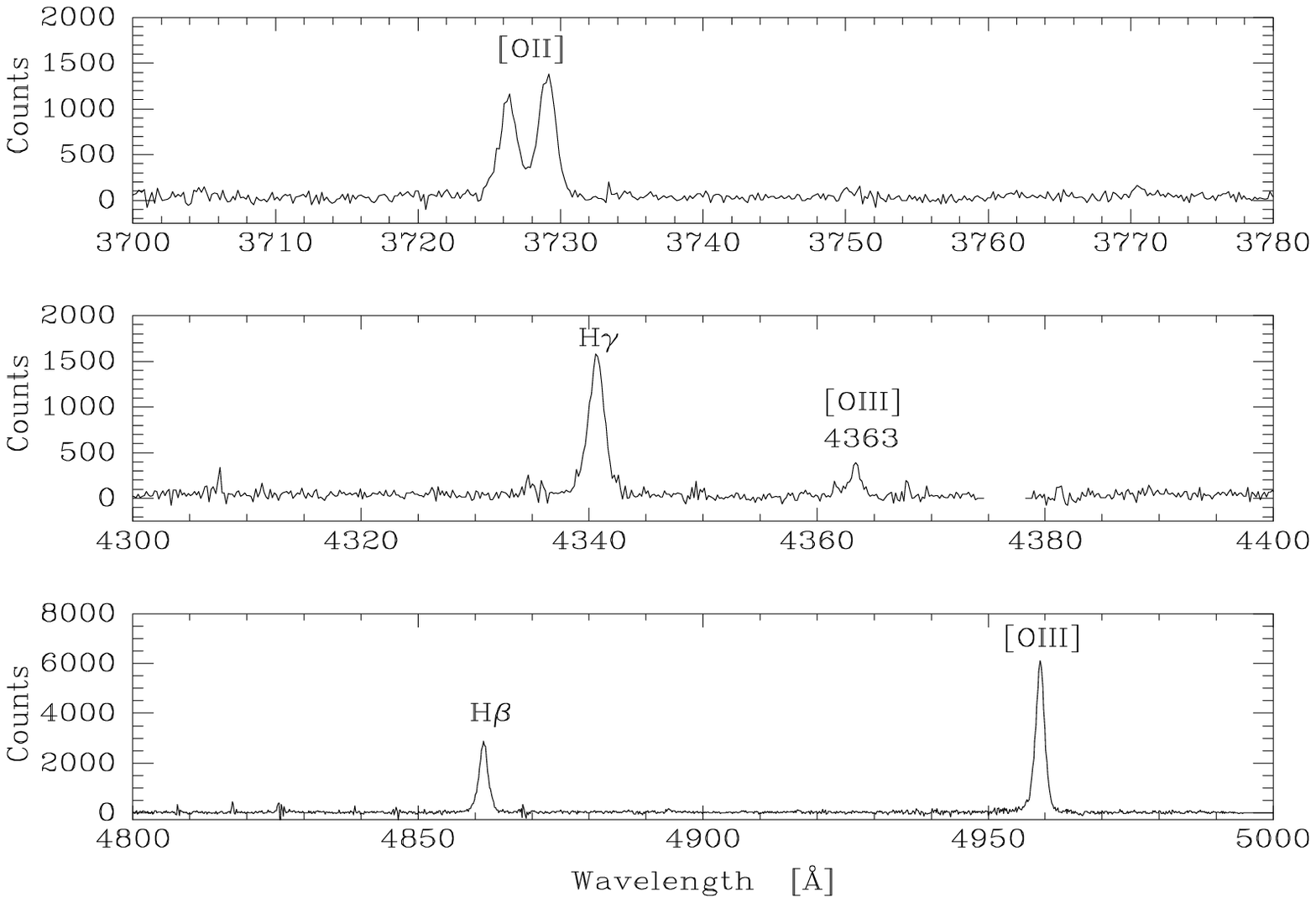} & \includegraphics[scale=0.50]{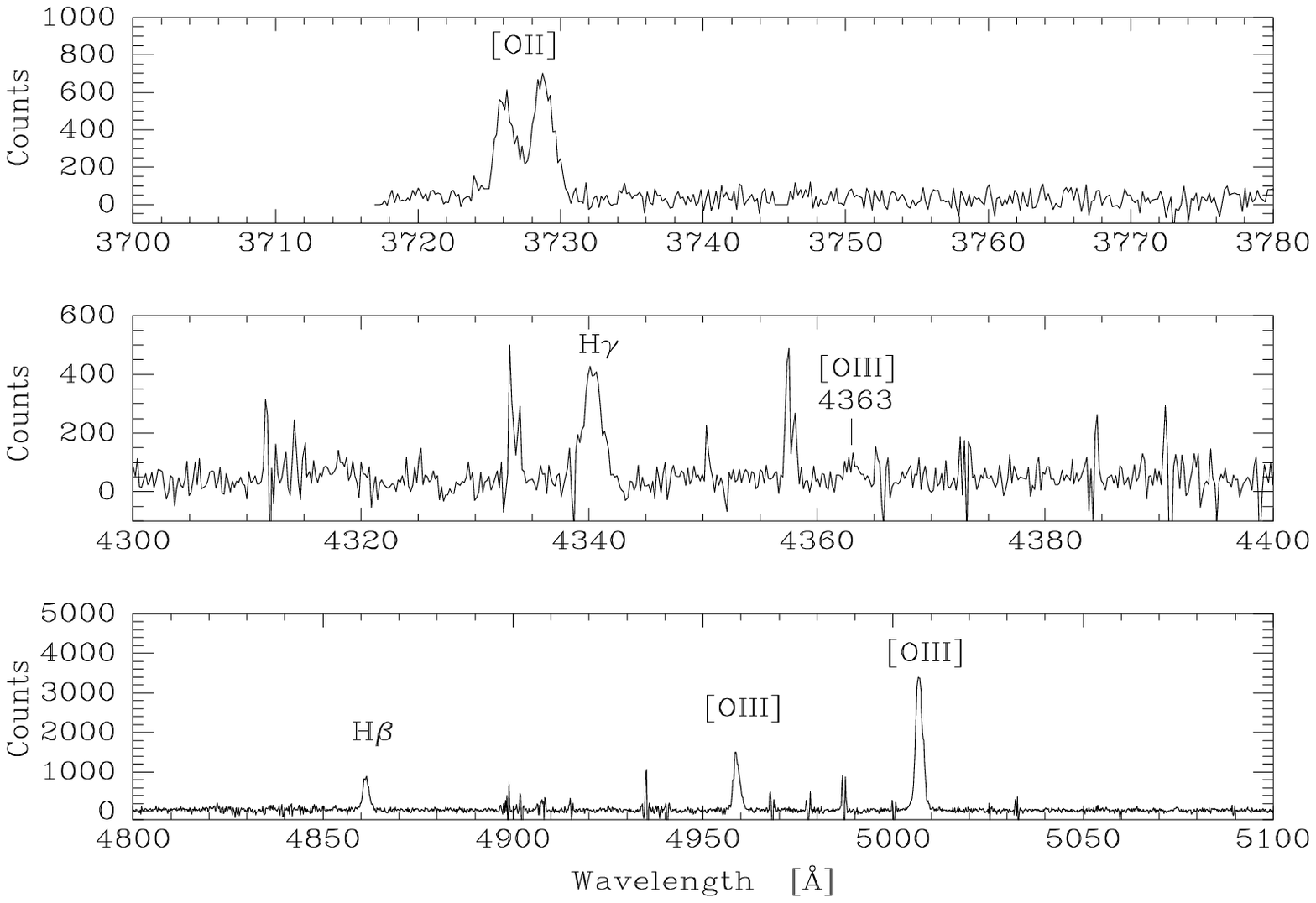} \\
\end{tabular}

\caption{Rest frame spectra of two metal-poor galaxies - DEEP2\_130016475 at $z=0.747$
  (left panel) and DEEP2\_31046514 at $z=0.789$ (right panel). Both show the temperature sensitive  
[\textsc{OIII}]$\lambda 4363$ line used to select the sample and the other oxygen 
emission lines used to measure the ionic and total abundances. Because
of the slit location on the mask, the spectrum in the left panel is
cutoff at a rest-wavelength blueward of [\textsc{OIII}]$\lambda5007$.}

\label{fig1}
\end{center}
\end{figure*}

We select galaxies showing the strong oxygen emission lines
[\textsc{OII}]$\lambda3727$, [\textsc{OIII}]$\lambda4959$ and that have
a reliable detection of the weak auroral line [\textsc{OIII}]$\lambda4363$
in their spectra. There is no need to include the [\textsc{OIII}]$\lambda5007$ in
the selection criteria since its intrinsic emission is linked to that of the [\textsc{OIII}]$\lambda4959$ line and
its reddening-corrected flux can be found as $I_{5007}=2.98 \times I_{4959}$ for normal star forming systems.
The combination of these requirements limit the redshift range to $0.69<z<0.88$, which includes 
$\sim 11000$ potential candidates.
This initial culling was made possible by the Weiner, B. J. \textit{priv. comm.} redshifts and
equivalent width measurements.
The remaining spectra were visually inspected
for reliable detections of the auroral [\textsc{OIII}]$\lambda4363$ line, and to
ensure the [\textsc{OII}]$\lambda3727$ doublet is cleanly
separated. This allows measuring the electron density accurately, without any previous assumptions.
We measured the  emission line parameters measured using the \textsc{IRAF} tasks \textsc{ngaussfit} and \textsc{splot}.
Our final sample consists of 22 sources, or about 0.2\% of the potential candidates in the relevant redshift window.
Figure \ref{fig1} shows two example spectra of the final sample.

The \emph{number count}\footnote{This error does not take into account flux calibration issues. These other error sources will be dealt with separately.}
$S/N$ ratio of the $H{\gamma}$ line is in the $15<S/N_{\gamma}<100$ interval with a typical value of $S/N_{\gamma}=50-60$.
In the case of the [\textsc{OIII}]$\lambda 4363$, we have $4.1<S/N_{4363}\leq 11.5$ with an average value of $S/N_{4363}=10.0$.
We thus have a worst-case uncertainty in the [\textsc{OIII}]$\lambda 4363 / \mathrm{H}\gamma$ of 25\%, with a typical value of 10\%.
The relative error in the [\textsc{OIII}]$\lambda 4363 / \mathrm{H}\gamma$ ratio 
is a major contributor to the error in the electron temperatures and
therefore oxygen abundances in the direct method we use to derive
metallicities. However, in this analysis, the determination of the
electron temperatures are robust relative to these uncertainties.

We stress here that, because of our visual sample selection, the
galaxies used here do not define a complete set in statistical terms. 
However, they are representative of the intermediate redshift
population of emission line galaxies with good metallicity
determinations that show [\textsc{OIII}]$\lambda 4363$.

The DEEP2 parent sample was selected from photometry obtained at the
Canada-France-Hawaii Telescope (CFHT) in the $B$,  $R$ and $I$ bands using the
CFHT 12k$\times$8k camera \citep{2004ApJ...617..765C}. The catalogues
were generated using the \textit{imcat} software
\citep{1995ApJ...449..460K} and the object magnitudes in $R$ are
measured within a circular aperture with a radius of 3 $r_g$ where $r_g$ is 
the optimal Gaussian profile unless  3 $r_g$ were less than 1$"$ , in
which case the flux is measured inside a 1$''$ aperture. The $B-R$ and $R-I$ colours
are measured with apertures of 1$''$ in order to minimise the noise \citep{2004ApJ...617..765C}.
The rest-frame magnitudes were obtained following \citet{cnaw06} and
use as templates a set of 34 local galaxies observed by \citet{1996ApJ...467...38K}. 
For each galaxy a parabolic fit between the synthetic $B-R$ and $R-I$ colours and $U-B$  measured at the galaxy redshift is used to estimate
the rest-frame B magnitude and $U-B$ colour. The \textit{rms} errors for the
K-corrections are usually smaller than 0.15 magnitudes
measured at the high redshift edge of DEEP2 ($z \sim 1.5$). The \textit{rms}
errors for the $U-B$ colours range from 0.12 mag at $z = 1.2$ (worst value)
to 0.03 mag at redshifts where the observed filters best overlap $U-B$
\citep{cnaw06}. For more detailed descriptions of both procedures we
refer the reader to \citet{2004ApJ...617..765C} and \citet{cnaw06}.

Table \ref{tab1} summarises the selected sample, where we present the
object identification from the DEEP2 catalogue, coordinates on the
sky, observed and rest-frame magnitudes, the rest-frame equivalent width of
H$\beta$ and the velocity dispersions. The rest-frame colours
are consistent with these sources containing vigorous star forming clusters.
The measured velocity dispersions indicate that none of the sources hosts Active Galactic
Nucleus (AGN) activity, as
defined by the criterion presented in \citet{ost89}, which sets the
limit between AGNs and normal star forming systems at around $180
\mathrm{km}\mathrm{s}^{-1}$. 

\begin{table*}
\caption{Full DEEP2 IDs, redshifts, J2000 coordinates, AB apparent and absolute
  B-band magnitudes, AB $UBV$ colours, $H\beta$ rest frame equivalent width and velocity dispersion.
Uncertainties in EW(H$\beta$) are a few \AA .}
\label{tab1}
\begin{tabular}{cccccccccc}
\hline 
ID & $z$ & RA & DEC & $m_{B}$ & $M_{B}$ & $U-B$ & $B-V$ & EW(H$\beta$) & $\sigma$  \\
   &     & $(hh:mm:ss)$  & $(dd:mm:ss)$ & $(mag_{ob})$ & $(mag)$   & $(mag)$ & $(mag)$  & ${(\AA)}$ & (km/s) \\
\hline 
\hline 
21007232 & 0.71659 & 16:47:26.19 & 02:19:00.81 & 23.40 & -20.04 & 0.47 & 0.35 & 36 & 37$\pm$2 \\
41022570 & 0.72138 & 02:27:30.46 & 00:02:04.43 & 23.43 & -19.47 & 0.39 & 0.31 & 310 & 29$\pm$2  \\
42009827 & 0.72939 & 02:29:33.65 & 00:01:44.53 & 23.97 & -19.32 & 0.36 & 0.25 & 80 & 30$\pm$2 \\
42025672 & 0.73143 & 02:29:02.03 & 00:02:00.54 & 22.87 & -20.07 & 0.36 & 0.28 & 122 & 48$\pm$3 \\
31047738 & 0.73232 & 23:26:41.18 & 00:01:13.06 & 22.60 & -20.53 & 0.33 & 0.23 & 93 & 38$\pm$3 \\
22006008 & 0.73280 & 16:51:08.82 & 02:19:04.23 & 24.52 & -18.92 & 0.44 & 0.32 & 13  & 25$\pm$5 \\
22032374 & 0.73839 & 16:53:06.12 & 02:19:57.79 & 23.60 & -19.94 & 0.51 & 0.38 & 38  & 35$\pm$2 \\
32018903 & 0.73961 & 23:30:55.46 & 00:00:47.86 & 23.73 & -19.65 & 0.48 & 0.36 & 89  & 46$\pm$7\\
13016475 & 0.74684 & 14:20:57.85 & 52:56:41.81 & 22.97 & -20.16 & 0.49 & 0.38 & 162 & 47$\pm$6 \\
22032252 & 0.74872 & 16:53:03.49 & 34:58:48.95 & 24.21 & -19.30 & 0.49 & 0.36 & 78 & 37$\pm$3 \\
31019555 & 0.75523 & 23:27:20.37 & 00:05:54.76 & 23.56 & -19.27 & 0.53 & 0.43 & 165 & 52$\pm$2 \\
14018918 & 0.77091 & 14:21:45.41 & 53:23:52.70 & 23.14 & -20.23 & 0.44 & 0.32 & 124 & 42$\pm$2 \\
12012181 & 0.77166 & 14:17:54.62 & 52:30:58.42 & 23.37 & -19.77 & 0.37 & 0.27 & 42 & 41$\pm$3 \\
41059446 & 0.77439 & 02:26:21.48 & 00:48:06.81 & 22.68 & -20.92 & 0.35 & 0.24 & 35 & 45$\pm$3 \\
41006773 & 0.78384 & 02:27:48.87 & 00:24:40.08 & 23.97 & -19.27 & 0.33 & 0.23 & 36 & 30$\pm$2 \\
31046514 & 0.78856 & 23:27:07.50 & 00:17:41.50 & 23.84 & -20.11 & 0.46 & 0.33 & 47 & 48$\pm$2 \\
22020856 & 0.79448 & 16:51:31.47 & 34:53:15.96 & 23.50 & -20.13 & 0.45 & 0.32 & 68 & 42$\pm$4  \\
22020749 & 0.79679 & 16:51:35.22 & 34:53:39.48 & 23.53 & -20.16 & 0.42 & 0.30 & 102 & 49$\pm$4 \\
22021909 & 0.79799 & 16:50:55.34 & 34:53:29.88 & 24.06 & -19.35 & 0.24 & 0.16 & 25 & 32$\pm$7   \\
21027858 & 0.84107 & 16:46:29.01 & 02:19:41.33 & 23.56 & -20.14 & 0.29 & 0.20 & 93 & 56$\pm$1 \\
22022835 & 0.84223 & 16:50:34.61 & 02:19:31.49 & 23.47 & -19.85 & 0.32 & 0.23 & 250 & 50$\pm$4 \\
31047144 & 0.85623 & 23:26:55.43 & 00:01:11.53 & 23.33 & -20.27 & 0.25 & 0.18 & 87 & 55$\pm$2 \\
\hline 
\end{tabular}
\end{table*}

\begin{figure*}
\begin{center}
\begin{tabular}{cc}
\includegraphics[scale=0.38]{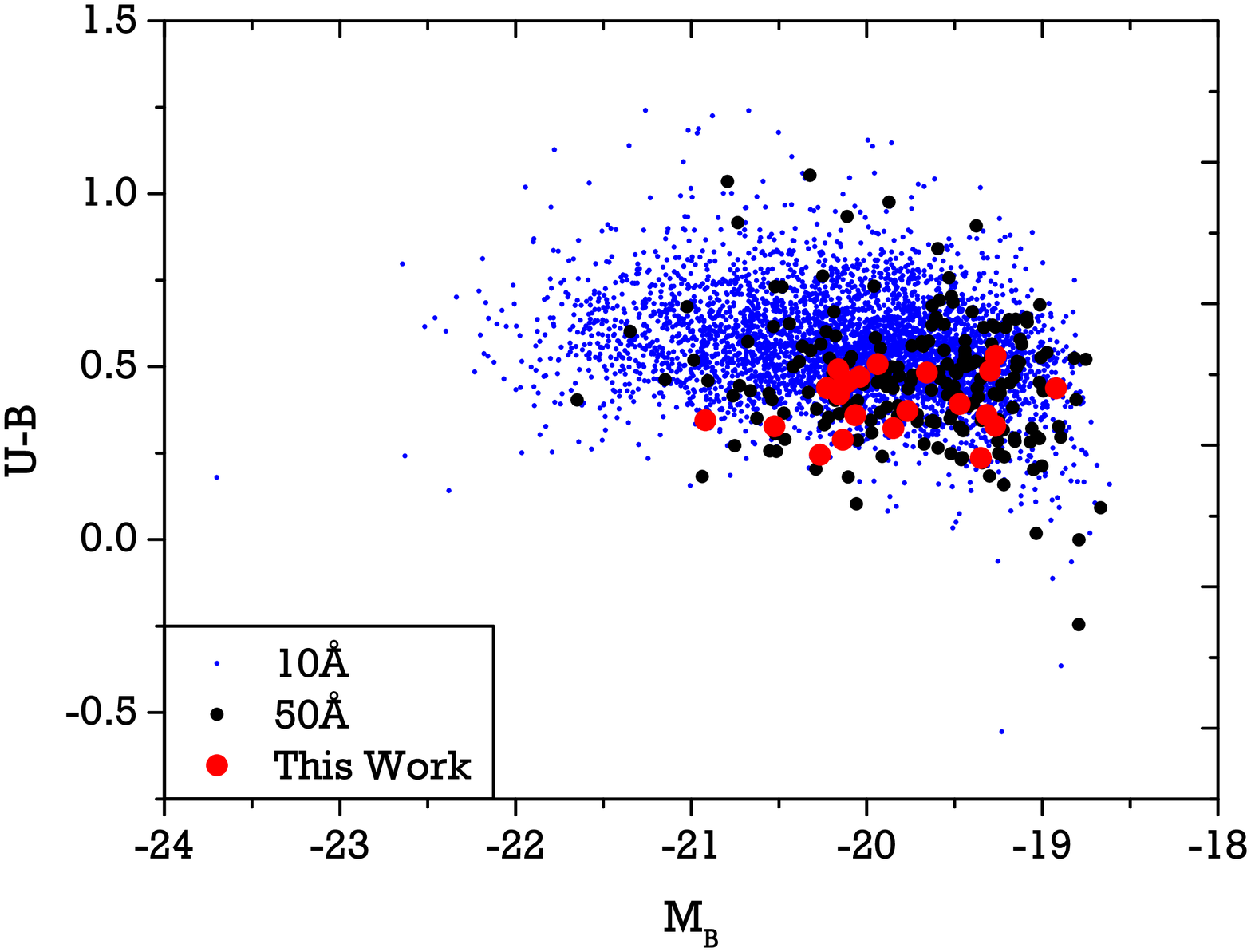} & \includegraphics[scale=0.38]{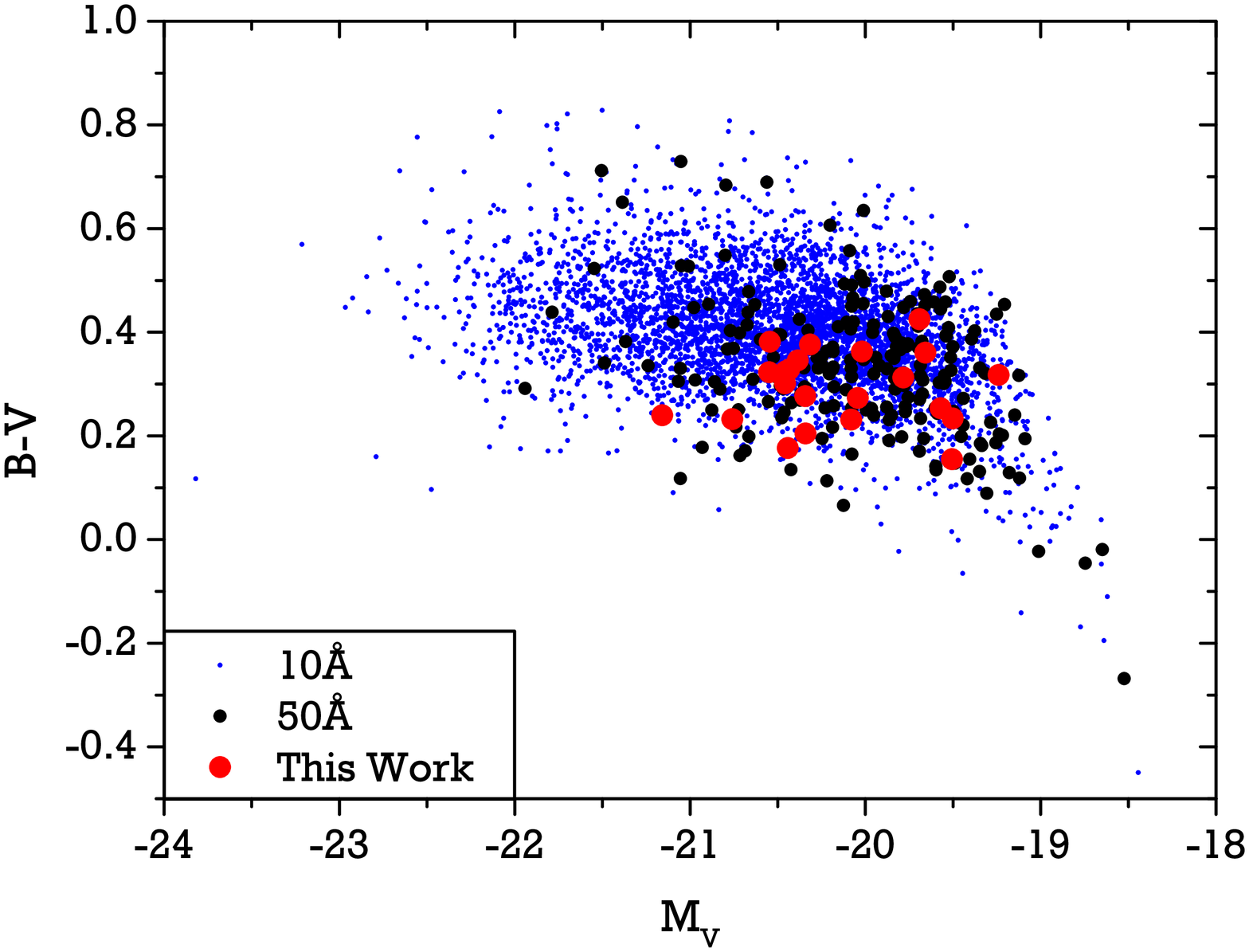} \\
\end{tabular}
\caption{$U-B$ \textsl{vs.} $B$ and $B-V$ \textsl{vs.} $V$ CMDs. The blue dots 
depict 4550 objects with $\mathrm{EW}(H\beta) \geq 10 \mbox{\AA}$. The black
dots show the 218 galaxies with $\mathrm{EW}H\beta) \geq 50\mbox{\AA}$.
Our sample is presented as large red solid circles.}
\label{fig2}
\end{center}
\end{figure*}

In Figure \ref{fig2}, we compare our selected sample against other DEEP2 sources in the same
redshift interval. The comparison galaxies were selected 
according to their $\mathrm{H}\beta$ line equivalent widths.
The purpose of this comparison is to highlight the nature of the [\textsc{OIII}]$\lambda 4363$ galaxies, showing that
their star formation episodes are both very intense and young.
The first comparison sample contains 4550 galaxies with $\mathrm{EW}_{\beta}>10\mathrm{\AA}$, which represents 
the general population of DEEP2 emission-line galaxies at these redshifts.
The second one contains 218 galaxies with $\mathrm{EW}_{\beta}>50\mathrm{\AA}$, which represents
either galaxies with an AGN or systems with very young and vigorous star forming episodes.  
It is seen that the [\textsc{OIII}]$\lambda 4363$ galaxies are very blue
and luminous. Some of the galaxies selected for this work are actually amongst
the bluest DEEP2 targets, even in the $U-B$ colour, hinting that our
spectroscopically selected sample must be comprised of galaxies with very intense starbursts.

\section{Results}
\label{resultaos}

\subsection{Relative flux calibration. Special case for the [\textsc{OII}]$\lambda 3727$ doublet.}
\label{rfc}

The DEEP2 spectra are not flux calibrated so that it is necessary
to make a relative flux calibration for each spectrum in order to
obtain the physical line ratios.
Here, we use the results provided by \citet{nkp_private}, who fits a smooth fourth order polynomial
to model the relative throughput with wavelength. This allows measuring meaningful observational line ratios, without the need to
perform a full flux-calibration of the spectra.
In our procedure, we first normalize the observed number counts of any given oxygen line to the
number counts of its nearest hydrogen Balmer line, obtaining an \emph{instrumental} line ratio.
Therefore, we first compute
$F(\mathrm{[OIII]}\lambda,\lambda 4959,5007)/F(\mathrm{H}\beta)$ and 
$F(\mathrm{[OIII]}\lambda 4363)/F(\mathrm{H}\gamma)$
in instrumental units and then apply the relative calibration
explained in \citet{nkp_private}, for which a brief summary is
presented as an appendix.  
In cases where two lines are close in
wavelength space, there is no need to apply the throughput correction, 
since the latter is almost identical for both the numerator and
denominator in the two expressions.
However, for the
[\textsc{OII}]$\lambda3727$ doublet the \citet{nkp_private} calibration is used
as there are no H series lines close enough in wavelength, as the case
of the [\textsc{OIII}] lines. In this case the relative flux
calibration uses the 
$F(\mathrm{[OII]}\lambda 3727) /F(\mathrm{H}\gamma)$ ratio.
The typical uncertainty in the empirical \citet{nkp_private}
calibration is of the order of 10\%, which must be propagated into the
aforementioned ratio. This is a major contributor to the error budget
in this specific line ratio, together with the uncertainty in fixing the
[\textsc{OII}]$\lambda 3727$ continuum level for vigorously star
forming objects.
An additional source of error is related to the possible presence
of an underlying stellar population contributing to the continuum, although
most of the objects studied here do not show strong absorption wings in the Balmer lines.
Typical values of the absorption $H\beta$ equivalent width found for
line-emiting star forming galaxies are in the range 0--3.5 \AA\ for
a spectral resolution of about 7 \AA\ \citep{1994ApJ...435..647I} .
For the higher spectral resolution of the data analysed here ($R \sim 5000$, at 7800\AA\ )
these contributions are considerably reduced, involving only the core of the line
\citep{1988MNRAS.231...57D}.
If we adopt a rest frame $\mathrm{H}\beta$ absorption equivalent width of the order of 1\AA\ ,
the underlying absorption correction is typically small, about 2\% in H$\beta$ and
3\% in H$\gamma$ due to the large Balmer emission equivalent widths.
However, given that we are measuring electron temperatures
via equation \ref{eq:ratio} below, the impact of this
uncertainty in our determination of the electron temperature
affects the $R_{O3}$ ratio defined below at the 1\% level at most.

\subsection{Temperatures, densities, and oxygen content calculations.}
\label{tdo}

Given the expected physical conditions of the line emitting regions in
the galaxies of this sample, it is safe to assume that all oxygen is either singly or
doubly ionised. The $O^{+}/O^{0}$ ratio is fixed to the
$H^{+}/H^{0}$ ratio via a charge exchange reaction, and is almost surely negligible
in this case. It would also require exceptionally hard
radiation in order to produce a significant amount of $O^{3+}$. 
Therefore, we use a two phase model with a low ionisation zone which
depends on the emission of the [\textsc{OII}]$\lambda 3727$ doublet, and
a high ionisation zone in which the \textsc{OIII} lines are formed.

Table \ref{tab2} summarises the required line ratios to solve the two-phase 
scenario, assuming the electron density of the higher ionization zone to be equal
to the electron density in the lower ionization zone. At any rate, the electron densities
found are well below the critical value for de-excitation. From our observations, we
can obtain $R_{O3}$ and $R_{ne}$. It is, however, not possible to calculate $R_{O2}$ as we cannot measure the
auroral [\textsc{OII}] lines as their observed wavelengths for this
sample of galaxies fall beyond the coverage of DEIMOS.

\begin{table}
\caption{Line ratios used to derive electron densities and temperatures. $I(nnnn)$ is the reddening-corrected intensity of each emission line.}
\label{tab2}
\begin{tabular}{ll}
Quantity. & Diagnostic. \\   \hline
$t_{e}$[\textsc{OIII}] & $R_{O3}=(I(4959)+I(5007))/I(4363)$ \\
$t_{e}$[\textsc{OII}] & $R_{O2}=I(3727)/(I(7319)+I(7330))$ \\
$n_{e}$[\textsc{OII}] & $R_{ne}=I(3726)/I(3729)$  \\ \hline 
\end{tabular}
\end{table}

We measure the physical ratio via the closest H line using the
normalised flux measurements defined in \S \ref{rfc} and simultaneously
account for internal extinction, assuming the theoretical case B recombination 
value for Balmer decrements I(H$\gamma$)/I(H$\beta$)=0.471, which
is the mean between the values corresponding to $T_{e}=10000K$ and
$T_{e}=20000K$ for $n_{e}=100 \mathrm{cm}^{-3}$, as is seen in \citet{ost89}.

\begin{equation} \label{eq:ratio}
R_{O3}=\frac{[f(4959)+f(5007)]/f(H\beta)}{f(4363)/f(H\gamma)}\frac{I(H\beta)}{I(H\gamma)}
\end{equation}
\noindent
where, in this case, $f(nnnn)$ represents the \emph{observed} number counts of each
emission line.
\indent

This expression is used because the $R_{O3}$ ratio involves lines with 
different wavelengths and therefore we need to minimise errors arising from the relative flux calibration. This is accomplished by measuring the 
line number counts in units of the nearest Balmer line.

On the other hand, the ratio between the two lines in the [\textsc{OII}]$\lambda3727$ doublet $R_{ne}$ bears the
lowest uncertainty, as it is not affected by flux calibration or reddening issues.
The error in our electron density determinations is thus dominated by the uncertainty in the electron
temperature of the O$^+$ zone.

We have derived the physical conditions of the $O^{++}$ dominated region
using the expressions given by \citet{guille08} for
the oxygen emission lines of \textsc{HII} galaxies. This formula is an approximation
for the statistical equilibrium model in a five level
atom. We present below the adequate fitting function they used from
the \textsc{temden} task of \textsc{iraf}, which is based on the program \textsc{fivel} (\citet{robertis87} and \citet{shaw95}):

\begin{equation}
t_{e}\mbox{[OIII]}=0.8254-0.0002415 \times R_{O3}+\frac{47.77}{R_{O3}}.
\end{equation}{\par}

The expected deviations in electron temperatures that could arise from the use of this expression are 5\% or lower.
In our case, the electron temperatures in the $O^{+}$ ionisation zone needs to be estimated from
$t_{e}\mathrm{[\textsc{OIII}]}$ because the [\textsc{OII}] auroral
lines are not included in the spectral range covered by the DEIMOS spectra.
Thus, the empirical expression given in \citet{pagel92}, which is based on a
fit to the theoretical models first presented in \citet{stasinska90} is used:

\begin{equation}
t_{e}^{-1}\mbox{[\textsc{OII}]}=0.5(t_{e}^{-1}\mbox{[\textsc{OIII}]}+0.8)
\end{equation}

The uncertainties in $t_{e}\mathrm{[\textsc{OII}]}$ are therefore a convolution between the $t_{e}\mathrm{[\textsc{OIII}]}$
errors and the variety of models\footnote{Chemical compositions, stellar atmospheres, ages, recombination case and geometry assumed, etc.} used to derive
the above expression for the electron temperature in the $O^{+}$ zone. These errors are bound to be higher 
than the errors for $t_{e}\mathrm{[\textsc{OIII}]}$, and are estimated
to be about 40\%. The errors quoted in Table \ref{tab3} only 
reflect the standard error propagation from the \citet{stasinska90} expressions and our  $t_{e}\mathrm{[\textsc{OIII}]}$ uncertainties. Once the 40\% intrinsic scatter
of the $t_{e}\mathrm{[\textsc{OII}]}$ calibration is taken into account, this translates into a 50\% error in the $O^{+}/H^{+}$ ionic abundance, or 0.3 dex.
Because most of the oxygen in the ionised gas-phase will probably be in the form of  $O^{2+}$,
the large uncertainties for the $O^{+}$ content will not
affect significantly the error budget for the final oxygen abundance.
Table \ref{tab3} shows the resulting electron temperatures and densities. It
is seen that the electron density is close to the expected value of
$100\mathrm{cm}^{-3}$. This is much lower than the critical density
$n_{\mathrm{crit}}$ where the higher energy level of the observed lines are as
likely to be de-excited by collisions as by radiative decay.
This indicates that our $t_{e}\mathrm{[\textsc{OII}]}$ calculations
are valid. For some galaxies, the resulting electron densities we
obtain are smaller than $50\mathrm{cm}^{-3}$, and for these only upper
limits are quoted.
Using the results above and the expressions of \citet{guille08} we
calculate the partial ionic abundances $12+\log (O^{+}/H^{+})$ and $12+\log (O^{++}/H^{+})$. These are then combined to obtain the
final oxygen abundance, where we assume the fractions of O$^0$ and O$^{3+}$ are negligible:

\begin{equation}
\frac{O}{H}=\left(\frac{O^{+}}{H^{+}}\right)+\left(\frac{O^{2+}}{H^{+}}\right)
\end{equation}

Table \ref{tab3} presents the resulting ionic and total abundances and the ratio between the two ionisation states $\log(O^{2+}/O^{+})$
in the gas. The errors are smaller than 0.1 dex for the oxygen
ionic and total abundances and between 0.1 and 0.15 dex for
the O$^{2+}$/O$^+$ ratio. The low oxygen abundances, corresponding to 
metallicities from 1/3 to 1/10 of the solar value, combined
with the extreme colour-magnitude properties of the sample
show that these galaxies are metal-poor.

\begin{table*}
\caption{Results from nebular analysis. Abundances, electron densities and electron temperatures for the O{2+} zone.}
\label{tab3}
\begin{tabular}{ccccccccc}
\hline 
ID  &  $z$  &  $n_{e}$  &  $T_{e} (O^{2+})$  &   $12+\log(O^{+}/H^{+})$  & $12+\log (O^{2+}/H^{+}$)  &  $12+\log(O/H)$  &  $\log(O^{2+}/O^{+})$  \\
    &       & $(cm^{-3})$  & $(10^{4} K)$ &  &  &  &  \\
\hline
  21007232  &   0.71659  &   100  &   1.16$\pm$0.09    &    7.72$\pm$0.11  &   7.98$\pm$0.10  &   8.17$\pm$0.11  &   0.26$\pm$0.21    \\
  41022570  &   0.72138  &   230  &   1.34$\pm$0.03    &    7.35$\pm$0.06  &   7.84$\pm$0.03  &   7.96$\pm$0.04  &   0.50$\pm$0.09   \\
  42009827  &   0.72946  &   70  &   1.47$\pm$0.07     &    7.42$\pm$0.08  &   7.80$\pm$0.05  &   7.95$\pm$0.06  &   0.38$\pm$0.13    \\
  42025672  &   0.73152  &   $<50$  &   1.17$\pm$0.03  &    7.42$\pm$0.07  &   8.03$\pm$0.04  &   8.13$\pm$0.04  &   0.61$\pm$0.10    \\
  31047738  &   0.73239  &   110  &   1.13$\pm$0.04    &    7.73$\pm$0.07  &   8.00$\pm$0.04  &   8.19$\pm$0.05  &   0.27$\pm$0.12  \\
  22006008  &   0.73282  &   100  &   1.78$\pm$0.13    &    7.45$\pm$0.09  &   7.49$\pm$0.07  &   7.77$\pm$0.08  &   0.04$\pm$0.16  \\
  22032374  &   0.73839  &   $<50$  &   1.53$\pm$0.09  &    7.29$\pm$0.08  &   7.69$\pm$0.07  &   7.83$\pm$0.07  &   0.40$\pm$0.15    \\
  32018903  &   0.73956  &   65  &   1.38$\pm$0.06     &    7.42$\pm$0.08  &   7.89$\pm$0.05  &   8.02$\pm$0.06  &   0.47$\pm$0.13    \\
  13016475  &   0.74684  &   200  &   1.29$\pm$0.02    &    7.08$\pm$0.06  &   7.97$\pm$0.02  &   8.03$\pm$0.02  &   0.89$\pm$0.08  \\
  22032252  &   0.74872  &   $<50$  & 1.61$\pm$0.08    &    7.14$\pm$0.08  &   7.60$\pm$0.06  &   7.73$\pm$0.06  &   0.46$\pm$0.13  \\
  31019555  &   0.75523  &   230  &  1.48$\pm$0.04     &    6.92$\pm$0.06  &   7.74$\pm$0.03  &   7.80$\pm$0.04  &   0.82$\pm$0.09   \\
  14018918  &   0.77091  &   180  &  1.18$\pm$0.04     &    7.51$\pm$0.07  &   8.04$\pm$0.05  &   8.15$\pm$0.05  &   0.53$\pm$0.12   \\
  12012181  &   0.77166  &   70  &   1.63$\pm$0.07     &    7.27$\pm$0.07  &   7.71$\pm$0.04  &   7.84$\pm$0.05  &   0.44$\pm$0.11  \\
  41059446  &   0.77439  &   110  &   1.44$\pm$0.09    &    7.28$\pm$0.08  &   7.74$\pm$0.07  &   7.87$\pm$0.07  &   0.46$\pm$0.15   \\
  41006773  &   0.78384  &   140  &   1.69$\pm$0.10    &    7.22$\pm$0.08  &   7.58$\pm$0.06  &   7.74$\pm$0.07  &   0.36$\pm$0.14   \\
  31046514  &   0.78856  &   80  &   1.24$\pm$0.05     &    7.48$\pm$0.08  &   7.93$\pm$0.06  &   8.06$\pm$0.06  &   0.45$\pm$0.14    \\
  22020856  &   0.79449  &   140  &   1.44$\pm$0.08    &    7.39$\pm$0.08  &   7.73$\pm$0.06  &   7.89$\pm$0.07  &   0.34$\pm$0.14    \\
  22020749  &   0.79679  &   80  &   1.69$\pm$0.12     &    7.32$\pm$0.08  &   7.45$\pm$0.07  &   7.69$\pm$0.08  &   0.13$\pm$0.16    \\
  22021909  &   0.79800  &   $<50$  &   1.61$\pm$0.05  &    7.14$\pm$0.06  &   7.70$\pm$0.03  &   7.81$\pm$0.04  &   0.56$\pm$0.10  \\
  21027858  &   0.84107  &   115  &   1.14$\pm$0.03    &    7.59$\pm$0.07  &   8.03$\pm$0.05  &   8.16$\pm$0.05  &   0.44$\pm$0.12   \\
  22022835  &   0.84220  &   370  &   1.60$\pm$0.20    &    7.19$\pm$0.12  &   7.62$\pm$0.13  &   7.76$\pm$0.13  &   0.43$\pm$0.25  \\
  31047144  &   0.85630  &   280  &   1.92$\pm$0.10    &    6.96$\pm$0.07  &   7.51$\pm$0.05  &   7.62$\pm$0.05  &   0.55$\pm$0.12   \\ \hline 
\end{tabular}

\end{table*}

\section{Discussion and Conclusions}
\label{discutir}

\begin{figure}
\begin{center}
\includegraphics[scale=0.37]{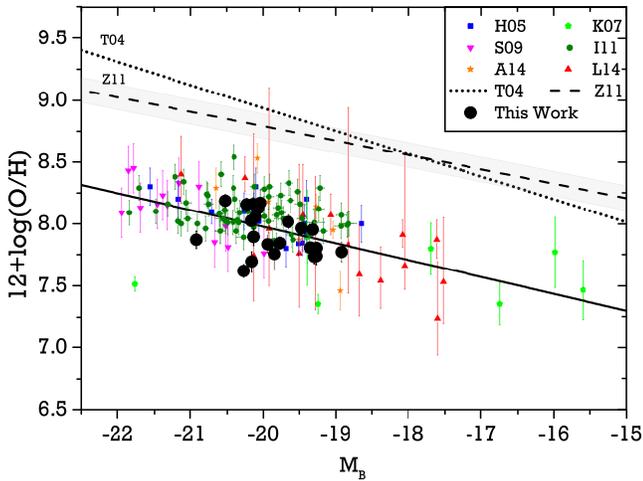}
\caption{Luminosity-Metallicity Diagram for intermediate star-forming galaxies for objects of this study and comparable samples from the literature. H05 includes 15 luminous star-forming galaxies from DEEP2. K07 includes 7 EELGs, 4 of them are extremely low metallicity galaxies. S09 includes 13 metal-poor galaxies. A14 includes 7 EELGs. L14 includes 16 metal-poor objects. I11 represents a sample of ``green peas" with good metallicity determinations. This work presents 22 new metal-poor galaxies. The dashed line represents the luminosity binned LZR by
\citet{zahid11} (Z11 in the diagram) with a $1-\sigma$ error area. The
dotted line on the upper zone of the diagram represents the Tremonti
luminosity-metallicity relation for local SDSS galaxies
(\citet{tremonti04}, T04 in the diagram) while the solid line
represents the best fit to the combined samples.}
\label{fig3}
\end{center}
\end{figure}

Metallicity is one of the most important parameters to understand
galaxy evolution and the total oxygen abundance is a common way to
trace the metallicities of line-emitting galaxies. 
The main result of this work is the direct metallicity determination using \textsc{[OIII]} electron temperatures, for a sample of 22
galaxies at intermediate redshifts ($0.69<z<0.88$). This is summarized in
the LZR diagram presented in Figure \ref{fig3}.
We measure total oxygen abundances between $1/10$ and $1/3$ of the solar value: 12+$\log$[O/H]$_{\odot}$ 
=8.69 \citep{asplund09} for approximately $-21<\mathrm{M}_{B}<-19$.
These results can be compared to \citet{hoyoskoo05} hereafter H05, \citet{kakazu07} hereafter K07, 
\citet{salzer09} hereafter S09, \citet{A14} hereafter A14 and
\citet{Ly14} hereafter L14.


The sample from K07 was collected from spectroscopic observations of 161 Ultra 
Strong Emission Line galaxies (USELs) using the DEIMOS spectrograph on the Keck II telescope. The galaxies are spread in a wide range of redshifts 
($0.38<z<0.83$) and its selection criteria are geared towards the detection of extremely low metallicities.
The majority of the objects in this sample have low luminosities and metallicities.
While no diagnostic diagrams could be used to flag objects hosting AGN 
or with the presence of shock heating, the high electron temperatures
measured for some galaxies ($20000<T(K)<30000$) suggest that some
objects have some contribution from these latter phenomena.

The sample from S09 presents some of the most luminous objects of this type
(-22 $<M_{B}<$ -20) at intermediate redshifts (0.35$<$z$<$0.41). The star-forming galaxies of 
this sample are selected from a wide-field Schmidt survey that picks 
emission line objects by the presence of H$\alpha$ emission in their objective-prism spectra. This naturally forces this sample to exclude low-luminosity objects.

The sample from A14 is composed by Extreme Emission Line Galaxies 
(EELGs) selected from the 20k zCOSMOS Bright Survey by their unusually large 
$\mathrm{[\textsc{OIII}]} \lambda5007$ equivalent widths. They are seven purely star-forming 
galaxies with redshifts from 0.43 to 0.63 and intermediate luminosities.

The sample from L14 encompasses a wide luminosity range (from $M_{B}$=-21.1 to 
-17.5) with abundances going from extreme metal poor galaxies 
(12+$\log$(O/H)$<$7.65) to galaxies with about half solar abundances and a redshift range similar to that of K07.
The data were obtained using optical spectroscopy with DEIMOS and the MMT Hectospec spectrographs.

The sample from H05 and this work have many features in common since both of
them have been taken from the DEEP2 redshift survey with the DEIMOS spectrograph. 
It covers the central region of the LZR diagram, though the presence of the [OII]$\lambda\lambda3726,3729$ doublet
within the wavalength range was not an explicit selection requirement
for H05. The redshift range of the H05 sample is $0.51<z<0.85$.

It is also possible to compare the sample presented here to
the ``green pea" population, first described by
\citet{2009MNRAS.399.1191C}. These systems were identified by
the Galaxy Zoo project because of 
their peculiar bright green colour and small sizes, being unresolved
in the Sloan Digital Sky Survey imaging. These galaxies show very strong [\textsc{OIII}]$\lambda  5007$ emission lines and very large H$\alpha$ equivalent widths up to 1000\AA .
Here, we have chosen for comparison the 66 $0.112<z<0.360$ ``green pea" sub-sample studied by \citet{2011ApJ...728..161I} (hereafter I11),  who collected a sample of 803 $0.02<z<0.63$ star-forming luminous compact galaxies. The global properties
of these star-forming luminous compact galaxies very closely resemble the properties of the ``green pea" population, but have been selected by both their spectroscopic and photometric signatures. These 66
``green peas" selected by \citet{2011ApJ...728..161I} were also
studied by  \citet{2009MNRAS.399.1191C}, but the former metallicities
were obtained using direct ($T_ {e}$) methods, which allows comparing them
to our sample in an optimal way. The oxygen abundances of the
\citet{2011ApJ...728..161I} ``green peas'' do not differ from those of
nearby low-metallicity blue compact dwarf galaxies. We here note that, at the resolution of SDSS, the objects in the sample presented here
would be almost point-like.


We find that our values for luminosities and metallicities are in good agreement with previous
determinations,  with our error estimates being smaller. Our sample shares the
same L-Z locus as the 66  ``green peas'' from I11, though our galaxies tend
to be less luminous and more metal-poor than the ``green peas''.

Taken together, all these studies define a locus in the LZR diagram,
that is offset from the local LZR of the SDSS sample of
\citet{tremonti04} towards lower metallicities.
These two loci can be represented by linear fits to the data, which
are shown in Figure \ref{fig3}; the local fit is represented a dotted line, and the intermediate redshift LZR
as a solid line. The linear regressions are:

\begin{eqnarray*}
12+\log(O/H) &=&-0.185(\pm0.001)M_{B}+5.238(\pm0.018) \\
12+\log(O/H )&=&-0.135(\pm0.019)M_{B}+5.272(\pm0.38) \\
\end{eqnarray*}
\noindent
for the local SDSS and intermediate redshift samples respectively.
\indent

Both lines have similar slopes, but are offset by a factor of 10 to lower oxygen content for the high luminosity end
considered in this work. In principle, this could be due to (i) the different nature of the objects involved, (ii) selection effects regarding both luminosity and chemical 
abundances, or (iii) a genuine evolutionary effect. However, we should
note that the abundances have been derived by two different methods for the local 
and for the distant samples. The local sample uses empirical
calibrations while our work uses direct abundance determinations.
The comparison of our $T_{e}$ abundances with those of the
intermediate $z$ sample of \citet{zahid11}, who used empirical
calibrations of the oxygen line equivalent widths, 
is a means of exploring the differences between both methods.
Here, it is critical that the parent samples in both works are essentially the same. The luminosity binned LZR, taking the median abundance value as metallicity, by
\citet{zahid11} is plotted in our Figure \ref{fig3}. It shows that, at the same luminosity, there is an $8-\sigma$
deviation between both abundance distributions. Given the numbers involved in the two samples (about 1700 \textsl{vs} 22), this large deviation
implies totally incompatible distributions. This huge discrepancy cannot be solved by taking into account the inherent uncertainties in both methods which amount to
0.2$dex$ at most. This could, in the best scenario, reduce the discrepancy to $4-\sigma$, which is still not plausible.

Our sample also shows similar luminosities, metallicities and optical appearances with the ``green peas" population of luminous star forming systems
found at lower redshift, although it tends to be fainter and less
metal rich than the ``green peas''.
The full description of the metallicity distribution
of star forming systems at intermediate redshift will require the use
of IR spectroscopy that allows detecting the [\textsc{SIII}] lines.
After carefully considering the systematics of the various metallicity determination methods, we conclude
that the metallicity distributions provided by other works like \cite{zahid11} and \cite{maier14}
and the one found in this work are not compatible.

\section*{Acknowledgments}

We are grateful to Ben Weiner, for providing us with the automated line
measurements.
Financial support has been provided by projects
AYA2010-21887-C04-03 (former Ministerio de Ciencia e In-
novaci\'on, Spain) and AYA2013-47742-C4-3-P (Ministerio
de Econom\'\i a y Competitividad), as well as the exchange
programme ‘Study of Emission-Line Galaxies with Integral-
Field Spectroscopy’ (SELGIFS, FP7-PEOPLE-2013-IRSES-
612701), funded by the EU through the IRSES scheme.
This work is based on observations taken at the W. M. Keck
Observatory, which is operated jointly by the National Aeronautics and
Space Administration (NASA), the University of California, and the
California Institute of Technology. Funding for the DEEP2 Galaxy
Redshift Survey has been provided by NSF grants AST-9509298,
AST-0071048, AST-0507428, and AST-0507483. 
We recognize and acknowledge the highly significant cultural role and
reverence that the summit of Mauna Kea has always had within the
indigenous Hawaiian community; it has been a privilege to be given the
opportunity to conduct observations from this mountain.

\appendix
\section{DEEP2 Relative flux calibration methods}

DEEP2 spectra need to be flux calibrated in order to measure meaningful observational line ratios. Here, we use the results provided by \citet{nkp_private} to perform a relative flux calibration of our sample's spectra. The aim of the relative flux calibration in our paper is to obtain the physical line ratios between oxygen and hydrogen lines by modelling the relative throughput of DEIMOS with wavelength.

Three methods were used to measure relative DEIMOS throughput, which is defined as the number of photons detected at the CCD over the total number of photons that leave the reference object. The basic algorithm to compute the throughput is the same in all the methods and it is the result of dividing a measured stellar spectrum by a standard spectrum associated with it.
The three methods used to compute a standard spectrum are: 1) Use a standard-star with well known SED. 2) Use the CFHT photometry to determine the corresponding stellar type in the Gunn-Stryker (G-S) catalog. 3)Use a least-square minimization to determine stellar type in the (G-S) catalog.

After carrying out an analysis of these methods, the throughput can be approximated by a fourth order polynomial:

$y({\lambda})= -77.9026+0.0395916\lambda-7.49911\times10^{-6}{\lambda}^{-2}$\\
$+6.29692\times10^{-10}{\lambda}^{3}-1.97967{\times}10^{-14}{\lambda}^{4} $ \\

where $\lambda$ is wavelength in \AA\ and y($\lambda$) is in units of throughput/\AA. The use of the previous calibration yields 10\% error in the emission line flux measurements.

\bsp
\label{lastpage}
\end{document}